\documentclass[pre,aps,amsmath,twocolumn,groupedaddress,notitlepage,longbibliography,superscriptaddress]{revtex4-1}

\usepackage{amssymb}
\usepackage{color}
\usepackage{MnSymbol}

\usepackage{tabularx}
\usepackage{psfrag}
\usepackage{graphicx}
\usepackage{amsmath}
\usepackage{natbib}
\usepackage{rotating}

\newcommand{\lbg}{l_{\mathrm{bg}}}  
\newcommand{\lr}{l_{\mathrm{r}}}
\newcommand{\lpro}{l_{\mathrm{pro}}}  
\newcommand{\lsel}{l_{\mathrm{sel}}}  
\newcommand{\erfc}{\mathrm{erfc}}
\newcommand{\schat}{\hat{s}_c}
\newcommand{\srhat}{\hat{s}_r}
\newcommand{\khat}{\hat{k}}
\newcommand{\Dshat}{\Delta \hat{s}}
\newcommand{\Dshatr}{\Delta \hat{s}_r}
\newcommand{\PRLcoop}{P^{\mathrm{coop}}_{RL}}
\newcommand{\PRLcomp}{P^{\mathrm{comp}}_{RL}}
\newcommand{\lstar}{l^{*}}
\newcommand{\urlf}[1]{\scriptsize{\texttt{#1}}}  
\renewcommand{\thefigure}{\arabic{figure}}
\renewcommand{\thetable}{\arabic{table}}

\renewcommand{\thesection}{\arabic{section}}
\renewcommand{\theequation}{\arabic{equation}}

\numberwithin{equation}{section}

\bibpunct{[}{]}{,}{n}{,}{,}
\begin{document}

\title{A model balancing cooperation and competition explains our right-handed world and the dominance of left-handed athletes}

\author{Daniel M. Abrams}
\email[E-mail address: ]{dmabrams@northwestern.edu}
\affiliation{Department of Engineering Sciences and Applied Mathematics, Northwestern University, Evanston, Illinois 60208, USA}
\author{Mark J. Panaggio}
\email[E-mail address: ]{markpanaggio2014@u.northwestern.edu}
\thanks{\newline M.J.P. contributed more to data collection and data analysis. D.M.A. and M.J.P. contributed equally to the model and its analysis.}
\affiliation{Department of Engineering Sciences and Applied Mathematics, Northwestern University, Evanston, Illinois 60208, USA}

\begin{abstract}
An overwhelming majority of humans are right-handed.   Numerous explanations for individual handedness have been proposed, but this population-level handedness remains puzzling.  Here we use a minimal mathematical model to explain this population-level hand preference as an evolved balance between cooperative and competitive pressures in human evolutionary history. We use selection of elite athletes as a test-bed for our evolutionary model and account for the surprising distribution of handedness in many professional sports.  Our model predicts strong lateralization in social species with limited combative interaction, and elucidates the rarity of compelling evidence for ``pawedness'' in the animal world. 

\centering\textbf{Keywords: }laterality; mathematical model; evolution; athletics; handedness

\end{abstract}

\maketitle

\section{Introduction}
\label{sec:Introduction}

Although the precise definition of handedness is often debated, it is widely accepted that roughly one in ten humans are left-handed \cite{raymond96,frayer11}.  Since prehistoric times, minor cultural and geographical variations in this percentage have been observed, but every historical population has shown the same strong bias toward right-handedness \cite{raymond96,frayer11,coren77,faurie05,llaurens09}. Both genetic and environmental factors seem to contribute to handedness for individuals \cite{llaurens09,brackenridge81,mcmanus91,francks07,uomini09}; nonetheless, individual handedness does not necessarily lead to species-level handedness.  It is well-established that for an individual, lateralization can be advantageous \cite{ghirlanda09,magat09}: for example, it allows for specialization of brain function \cite{mcmanus91,levy77} which may lead to enhanced cognition through parallel information processing \cite{rogers04,vallortigara05}. At the species level, however, the advantage of lateralization is not well understood \cite{vallortigara05}.  Negative frequency-dependent selection alone, a primary mechanism by which polymorphisms are maintained \cite{ayala74,billiard05}, can only produce a balanced distribution of left- and right-handers due to the symmetry inherent in handedness.

\section{Our Model}
\label{sec:Our Model}
There have been various attempts to account for species-level asymmetry with the use of ``fitness functions'' \cite{ghirlanda09,billiard05}.  We propose a different approach to the problem. We define a function $P_{RL}(l)$ representing the mean probability that a right-handed individual (male or female) bears left-handed offspring in a given time period.  A minimal model for the evolution of the societal fraction left-handed $l$ in terms of this arbitrary frequency-dependent transition rate $P_{RL}(l)$ is given by
\begin{equation}
\label{ode_v1}
\frac{dl}{dt} = (1-l) P_{RL}(l)-l P_{LR}(l).
\end{equation}

We assume symmetry between right- and left-handers (supplementary material section \ref{sec:The symmetry of handedness}), so that we may write $P_{LR}(l) = P_{RL}(1-l)$ to obtain
\begin{equation}
\label{ode_v2}
\frac{dl}{dt} = (1-l) P_{RL}(l)-l P_{RL}(1-l).
\end{equation}
This function incorporates frequency-dependent selection effects, and can be approximated given a biological model for inheritance (supplementary material section \ref{sec:Phenotypic model for population dynamics}, figure \ref{fig:ldotplot}).

In a purely competitive society, it would be natural to assume $\PRLcomp$ to be a monotonically decreasing function of $l$.  When left-handers are scarce, they have an advantage in physical confrontations due to their greater experience against right-handers and the right-handers' lack of experience against them.  As their numbers grow, that advantage weakens \cite{faurie05,billiard05}.

However, in a purely cooperative society, physical confrontations would not exist, and all individuals would tend to the same handedness to increase cooperative efficiency and eliminate the fitness disadvantage of the minority handedness \cite{faurie05,ghirlanda09,vallortigara05}. (The modern presence of a higher accidental death rate for left-handers \cite{halpern88,coren91,aggleton93} demonstrates that a fitness differential persists today.)  Thus $\PRLcoop$  would increase monotonically with $l$.

For a system involving both cooperative and competitive interactions, we therefore write \[P_{RL}(l) = c\PRLcoop(l) + (1-c)\PRLcomp(l),\] where $0 \leq c \leq 1$ represents the degree of cooperation in interactions and the monotonicity properties of each component function are as given above. For physically reasonable choices of these functions, there may exist one, three, or five fixed points $\lstar$ in this system, depending on the value of $c$.

\section{Analysis of Model}
\label{sec:Analysis of Model}

\begin{figure}[h] 
\begin{center}  
\includegraphics[width=84mm]{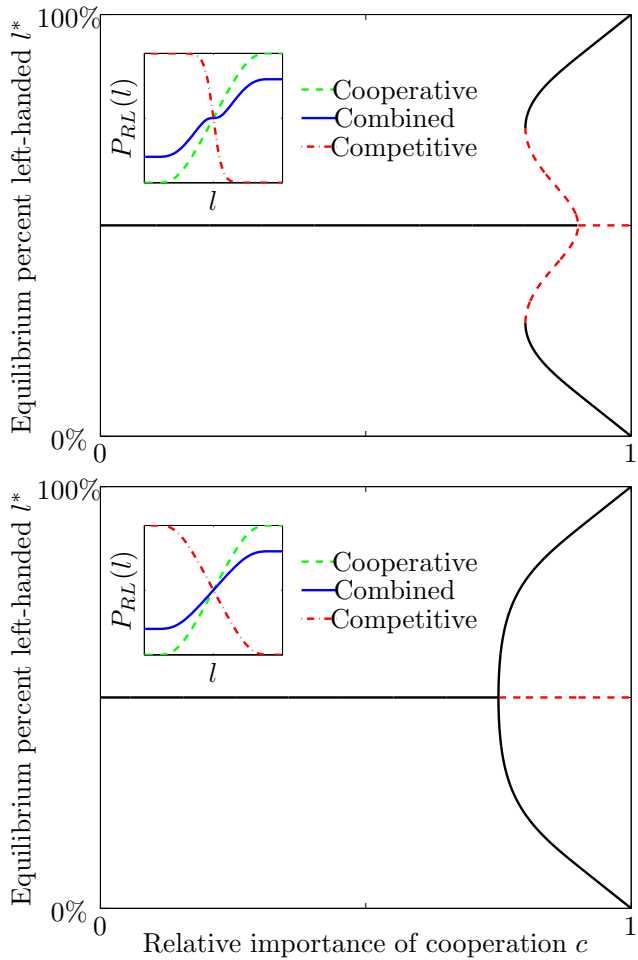}
\caption{Equilibrium percentages left-handed as a function of societal cooperativity.  Solid lines indicate stable equilibria, dashed unstable.  (a) Monotonic $\textrm{d}P_{RL}/\textrm{d}t$ on $(0,1/2)$. (b) Nonmonotonic $\textrm{d}P_{RL}/\textrm{d}t$ on $(0, 1/2)$.  Insets:  $P_{RL}$ and its component functions  $\PRLcoop$, $\PRLcomp$.}
\label{fig1}
\end{center}
\end{figure}

Figure \ref{fig1} shows the typical positions of stable and unstable equilibria for equation \eqref{ode_v1}, where $\PRLcoop$  and $\PRLcomp$  have been chosen to be generic sigmoid functions (sigmoid functions arise naturally in models where separate fitness functions exist for left- and right-handers --- see supplementary material section \ref{sec:Relating fitness functions to probabilistic transition rates}). When the degree of cooperation $c$ is less than a critical threshold, the only stable equilibrium is $\lstar = 0.5$: a 50/50 split between left-hand and right-hand dominant individuals.  This is consistent with studies showing individual but not population-level bias in various species \cite{uomini09,rogers02}.

When the degree of cooperation exceeds a critical threshold, two new stable equilibria appear as a result of either a subcritical or supercritical pitchfork bifurcation (depending on the exact form of the function $P_{RL}$).  These equilibria indicate population-level lateralization as seen in human society.  The fraction right- or left-handed will depend on the exact value of the cooperation parameter $c$.

There is a qualitative difference between the two situations depicted in figure \ref{fig1}, a difference which holds for a broad class of sigmoid functions $P_{RL}$.  In the case of the subcritical pitchfork (figure \ref{fig1}a), no weak population lateralization should ever be observed since equilibria near 50\% are unstable; however, in the case of the supercritical pitchfork (figure \ref{fig1}b), population lateralization near 50\% will be possible, though only stable for a small range of values of $c$.  Both suggest that weak population lateralization (fractions {\raise.17ex\hbox{$\scriptstyle\sim$}}$50\%-70\%$) should be rare in the natural world, while indicating that a high degree of cooperation may be responsible for the strong lateralization (fractions {\raise.17ex\hbox{$\scriptstyle\sim$}}$70\%-100\%$) observed in some social animals (e.g., humans, parrots \cite{harris89}).

\section{Comparison of Model Predictions with Data}
\label{sec:Comparison of Model Predictions with Data}

Thus far we have attempted to describe the evolution of the fraction left-handed in populations of lateralized individuals.  Ideally, we would compare predicted equilibria of equation \eqref{ode_v2} to data from animal populations exhibiting varying degrees of cooperation.  For most species, however, quantifying the degree of cooperation is difficult, and data on population-level lateralization is scarce and sometimes contradictory.  This lack of information about the natural world leads us to examine the proxy situation of athletics, where data on handedness and cooperation is more easily accessible.

To explain the observed fraction of athletes left-handed, it is important to model the selection process because athletics, unlike evolution, should not cause changes in the population's background rate of laterality.  We treat athletic skill s as a normally distributed random variable, and assume that minority handedness creates a frequency-dependent shift $\Delta s$ that modifies the randomly distributed skill.  We then model an ideal selection process as choosing the $n$ most skilled players from a population of $N$ interested individuals.  Such a model (derived in detail in supplementary material section \ref{sec:Derivation of athletic selection model}) predicts that the professional fraction left-handed $\lpro$ will depend on the fraction selected $\psi=n/N$, and is determined implicitly by the equation
\begin{equation}
\label{lpro_eq}
\lpro = \frac{\lbg}{2} \erfc(\schat-\Dshat)/\psi,
\end{equation}
where $\lbg \approx 10\%$ is the background rate of left-handedness, erfc is the complementary error function, $\schat$ is the normalized cut-off in skill level for selection, and $\Dshat \propto \lstar-\lpro$ is the normalized skill advantage for left-handers.  Here $\lstar$ represents the fraction of the population that would be left-handed in a world consisting only of interactions through the sport under consideration.  Its value is determined from equation \eqref{ode_v2}, with a choice of parameter $c$ appropriate for the sport under consideration ($P_{RL}$ is reinterpreted as the mean probability that a right-handed player is replaced by a left-hander in a given time period).  Note that $\lpro$ must lie between $\lbg$ and $\lstar$: with very high selectivity $(\psi\rightarrow0)$  equation \eqref{lpro_eq} implies that $\lpro\rightarrow\lstar$ , and with very low selectivity $(\psi\rightarrow1)$ $\lpro \rightarrow \lbg$. 

\begin{figure}[h] 
\begin{center}  
\includegraphics[width=84mm]{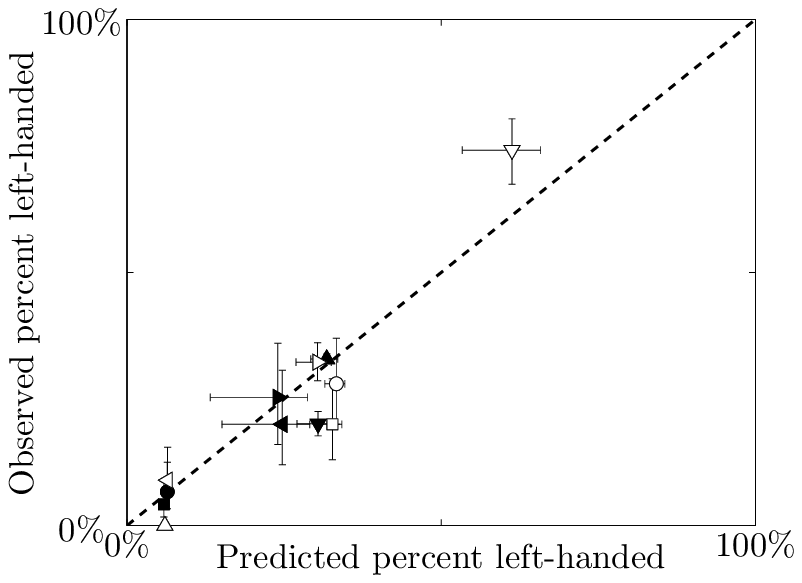}
\caption{Observed percentage left-handed versus predicted for professional athletes of various sports. $\filledmedtriangleup=$Baseball (MLB), $\filledmedtriangledown=$Boxing (Men's), $\filledmedtriangleleft=$Fencing (Men's), $\filledmedtriangleright=$Fencing (Women's), $\bullet=$Football (Quarterbacks, NFL),  $\filledmedsquare=$Golf (PGA),  $\medtriangleup=$Golf (LPGA),  $\medtriangledown=$Hockey (Right Wings, NHL), $\medtriangleleft=$Hockey (Left Wings, NHL) , $\medtriangleright=$Hockey (Other, NHL),  $\circ=$Table Tennis (Men's), $\medsquare=$Table Tennis (Women's). Dashed line represents perfect agreement between predicted and observed values. Vertical error bars correspond to 95\% confidence intervals ($p = 0.05$); horizontal error bars correspond to predictions using plus or minus one order of magnitude in $N$, the primary source of uncertainty.  Left-handed advantage $\Dshat=\khat(\lstar-\lpro)$ where $\khat = 1.6108$ and both $\lstar$ and $\lpro$ vary from sport to sport (see table \ref{datatable} and supplementary material section \ref{sec:Data summary})}
\label{fig2}
\end{center}
\end{figure}

Figure \ref{fig2} shows our application of equation \eqref{lpro_eq} to various professional sports.  To reduce arbitrary free parameters, we assume that the cooperativity $c$ is close to zero for physically competitive sports and one for sports (e.g., golf) that require lateralized equipment or strategy.  Figure \ref{fig1} then implies that the ideal equilibrium fraction left-handed $\lstar$ will be 50\% when $c$ is close to zero, and will be either 0\% or 100\% when $c=1$.

The predictions for figure \ref{fig2} were made by varying a single free parameter $k$, the constant of proportionality for the frequency-dependent skill advantage $\Dshat=k(\lstar-\lpro)$.  To avoid over-fitting, we took this to be a constant across all sports; given sufficient data, different values of $k$ could be estimated independently for each sport.  The fraction selected $\psi$ was estimated from the ratio of professional athletes to the number of frequent participants for each sport (see supplementary material section \ref{sec:Data summary} for details).

For the sport of baseball, the great abundance of historical statistical information allows us to validate our proposed selection mechanism.  To do so, we use our model to predict the cumulative fraction left-handed $\lr$ as a function of rank $r$, then compare to data.

In sports where highly-rated players interact with other highly-rated players preferentially (e.g., boxing), we expect the left-handed advantage $\Dshatr\propto\lstar-\lr$ to be rank-dependent (i.e., depending on the fraction left-handed at rank $r$).  However, within professional baseball leagues, all players interact with all other players at nearly the same rate, so the left-handed advantage $\Dshatr=\Dshat$ should be independent of rank, i.e., a constant.  This leads us (see supplementary material section \ref{sec:Derivation of athletic selection model} for derivation) to the equation
\begin{equation}
\label{lr_eq}
\lr=\frac{\lbg}{2} \erfc(\srhat-\Dshat)/(r/N).
\end{equation}

\begin{figure}[h] 
\begin{center}  
\includegraphics[width=84mm]{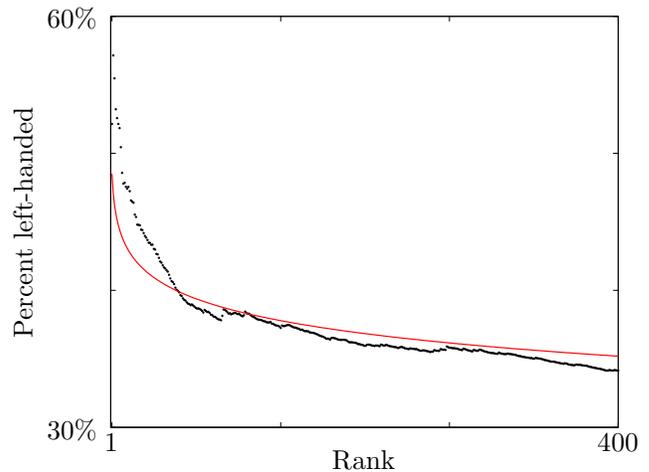}
\caption{Cumulative fraction left-handed versus rank for seasonal top hitters in baseball, 1871-2009.  Thin red line represents theoretical prediction from equation \eqref{lr_eq}.  Each black data point at rank $r$ represents the left-handed fraction of all U.S. born players that finished a season ranked in the top $r$ by total hits.  The left-handed advantage $\Dshat=0.3003$ was computed by finding the least-squares best fit.  This value differs slightly from the value used for baseball in figure \ref{fig2} ($\Dshat=0.2755$) suggesting that, in practice, the proportionality constant $k$ may vary from sport to sport.  }
\label{fig3}
\end{center}
\end{figure}

Figure \ref{fig3} shows the predictions of equation \eqref{lr_eq} as applied to the top-ranked baseball players from 1871 to 2009.  Only one free parameter was varied: the left-handed advantage $\Dshat$. All other parameters were constrained by known data \cite{lahman11}.  The surprisingly good fit to this nontrivial curve can be seen as supporting evidence for the selection model.  Together with the accuracy of predictions in figure \ref{fig2}, this supports the conclusion that the equilibria of equation \eqref{ode_v2} are indeed relevant to real-world lateralized systems.

\section{Discussion}
\label{sec:Discussion}

Despite the good agreement of our predictions with real-world data, we acknowledge that there are limitations in reducing a complex adaptive system to a simple mathematical model. Our model includes undetermined functions that would be difficult to measure precisely (although we found that qualitative predictions are robust --- see supplementary material section \ref{sec:Parameter sensitivity analysis for probabilistic model}).  However, they can be roughly approximated from available data and may be easier to estimate than fitness functions proposed in other models.  Sports data may not be completely analogous to data from the natural world; hence, quantitative analysis of lateralization in social animal groups may be a fruitful line of future research.

Given the limited data on population-level lateral bias in the natural world, we feel that analysis of athletics provides new insight into the evolutionary origins of handedness.  Our model predictions match the observed distribution of handedness in baseball with just a single free parameter.  When applied to 12 groups of elite athletes, the same model does a good job of estimating the fraction left-handed in each, suggesting that the proposed balance between cooperation and competition accurately predicts the ideal equilibrium distribution of handedness. Our model is general enough to be applied to any species of animal, and may also have use in understanding population-level lateralized adaptations other than handedness, both physical and behavioural.

\section{Conclusions}
\label{sec:Conclusions}

The model we have presented is the first to take a dynamical systems approach to the problem of laterality.  It allows for the prediction of conditions under which population-level lateral bias can be expected to emerge in the animal world and its evolution over time. We exploit the connection between natural selection and selection in professional sports by introducing a novel data set on handedness among athletes, demonstrating a clear relationship between cooperative social behaviour and population-level lateral bias.

\begin{acknowledgements}
This work was funded by Northwestern University and The James S. McDonnell Foundation.  The authors thank R. N. Gutenkunst and R. J. Wiener for useful correspondence.
\end{acknowledgements}



\onecolumngrid
\newpage
\section*{Supplementary Material}
\appendix
\renewcommand{\thefigure}{S\arabic{figure}}
\renewcommand{\thetable}{S\arabic{table}}
\renewcommand{\thesection}{S\arabic{section}}
\renewcommand{\theequation}{S\arabic{equation}}
\renewcommand{\appendixname}{}
\setcounter{secnumdepth}{1}  
\setcounter{section}{0}
\setcounter{equation}{0}
\setcounter{figure}{0}
\numberwithin{equation}{section}

\section{The symmetry of handedness}
\label{sec:The symmetry of handedness}

There have been various attempts to account for species-level laterality.  Billiard et al.~\cite{billiard05}~point out that left-handedness is ``associated with several fitness costs.''  Thus the population-level bias can be maintained through a balance between a frequency-dependent fitness function and a constant fitness cost.  An alternate model by Ghirlanda et~al.~\cite{ghirlanda09}~suggests that the combination of ``antagonistic'' and ``synergistic'' interactions and their associated frequency-dependent fitness functions can create an evolutionarily stable equilibrium with an asymmetric (and non-trivial) distribution of lateralization.

While both models have merit, they disagree on a fundamental question: Are left- and right-handedness interchangeable? In other words, is a mirror image world of 90\% left-handers and 10\% right-handers equally plausible?  Billiard et al.~suggest that the fitness costs ``such as lower height and reduced longevity...are not likely to be frequency-dependent.'' Thus these fitness costs break the symmetry and guarantee that only the observed handedness distribution is possible. 

However, the aforementioned fitness costs have only been observed in a biased population consisting of lateralized individuals. For example, Aggleton et al.~\cite{aggleton93}~show that left-handers are more likely to die prematurely, and that this effect is at least partially due to ``increased vulnerability to both accidental death and death during warfare.''  They go on to argue that ``the most likely explanation for the increase in accidental death among the left-handed men concerns their need to cope in a world full of right-handed tools, machines, and instruments.''  Clearly if left-handedness were more prevalent than right-handedness, then left-handed tools would also be more common, and, as a result, right-handers instead would experience increased risk of accidental death. Thus, it is likely that these fitness costs are frequency-dependent and symmetric.

In contrast to Billiard et al., Ghirlanda's model assumes that left- and right-handedness are indeed interchangeable.  Given an initial distribution of 50\% left-handers and 50\% right-handers, this model predicts that both the observed distribution and its mirror image are equally likely equilibrium outcomes.  On this point, our probabilistic model agrees with Ghirlanda et al.  Given the fact that there is no reason to expect that right-handedness is inherently superior to left-handedness from a fitness perspective, we assume that the probabilistic transition rates satisfy the symmetry condition $P_{RL}(l)=P_{LR}(1-l)$.

\section{Phenotypic model for population dynamics}
\label{sec:Phenotypic model for population dynamics}

The probabilistic model described in the main text provides a description of the population dynamics from a top-down perspective.  However, it may be more intuitive to consider the dynamics from a bottom-up perspective.  Here we develop a model using reproductive fitness arguments on the level of individuals and show that this produces essentially the same result.     

\vspace{0.5cm}
\noindent \textbf{Iterative Model}

Let us define $N$ to be the total population size, and $L$ and $R$ to be the number of left- and right-handers, respectively.  We make the simplifying assumption that $L+R=N$, i.e., there are no ambidextrous individuals.  Also, define handedness fractions $r=R/N$ and $l=L/N$ so that $r + l = 1$.

Suppose that in this population, individuals repeatedly pair off and reproduce, adding new individuals to the population in each generation.  From an evolutionary perspective, the expected number of offspring that individuals produce should be dependent on their fitness.  With all other factors being equal, an individual's fitness should be determined by his or her handedness and the distribution of handedness in the population.  Thus we define $b_R(l)$ and $b_L(l)$ to be expected number of offspring born to right- and left-handers. Also, suppose that a pairing $XY$ produces left-handed offspring with probability  $\sigma_{XY}$ and right-handed offspring with probability $1-\sigma_{XY}$, where $X$ and $Y$ represent the dominant hands of the parents.  We expect $\sigma_{XY}=\sigma_{XY}(l)$ to be a frequency dependent function. 

Thus there are 6 possible reproductive interactions (we ignore gender effects here for simplicity---including them should not qualitatively change the results):  
\begin{align*}
R+R &\xrightarrow{\sigma_{RR}} L & R+L &\xrightarrow{\sigma_{RL}} L &L+L &\xrightarrow{\sigma_{LL}} L \\
R+R &\xrightarrow{1-\sigma_{RR}} R & R+L &\xrightarrow{1-\sigma_{RL}} R & L+L &\xrightarrow{1-\sigma_{LL}} R ~.
\end{align*}
At each iteration, we suppose that a fraction $D$ of current members dies off. Then the number of left- and right-handers in the new generation is: 
\begin{align*}
L_{n+1} &=\left[\frac{b_R\sigma_{RR}R_n^2+(b_R+b_L)\sigma_{RL}R_nL_n+b_L\sigma_{LL}L_n^2}{N_n}\right]+\left[1-D\right]L_n \\
R_{n+1} &=\left[\frac{b_R(1-\sigma_{RR})R_n^2+(b_R+b_L)(1-\sigma_{RL})R_nL_n+b_L(1-\sigma_{LL})L_n^2}{N_n}\right]+\left[1-D\right]R_n ~.
\end{align*}

\vspace{0.5cm}
\noindent \textbf{Continuous Model}

The iterative perspective is intuitive but has limited predictive capacity.  One limitation is that the number of left- and right-handers are only defined at fixed intervals.  To remove this obstacle, we transform the discrete model into a continuous model.  We set $\beta_X(l)=\frac{b_X(l)}{\Delta t}$ to be the instantaneous birth rate for individuals with handedness $X$ and $\delta=\frac{D}{\Delta t}$ to be the instantaneous death rate. We set $X(t)=X_n$ and $X(t+\Delta t)=X_{n+1}$, and let $\Delta t\rightarrow 0$ to obtain ordinary differential equations for the evolution of $R(t)$,  $L(t)$ and $N(t)$.  However, these equations are not independent.  In fact, we are interested only in the the evolution of $l(t)$, which is governed by     
\begin{equation}  \label{phenotypic_model}
  \frac{dl(t)}{dt} = \left[ \beta_R\sigma_{RR}r(t)^2 + (\beta_R+\beta_L)\sigma_{RL}r(t)l(t) + \beta_L\sigma_{LL}l(t)^2 \right] - \beta_{\textrm{eff}} l(t)~.  
\end{equation}

In order to analyse this ODE assumptions about the functions $\beta_X(l)$ are needed:  a first order assumption is that these should be linear functions of the frequency $l$, and by symmetry, we expect that $\beta_L(l(t))=\beta_R(1-l(t))$.
According to Aggleton et al.~\cite{aggleton93}, the average lifespan of right-handers is 3.31\% longer than their left-handed counterparts, with much of the difference attributable to higher rates of premature death in war and accidents.  This indicates that in a society consisting of roughly 90\% right-handers, left-handers appear to have a lower fitness.  This fitness differential should be reflected in the model's reproductive rates. The overall birth rate $\beta_{\textrm{eff}}$ in the U.S., a weighted average of $\beta_R$ and $\beta_L$, is known: $\beta_{\textrm{eff}}(l^*)=0.01383$ \cite{worldfactbook}.   If we assume that $\beta_R\left(l^*\right)=1.0331\cdot\beta_L\left(l^*\right)$ then linearity and symmetry allow us to derive expressions for $\beta_R$ and $\beta_L$:
\begin{align*} 
  \beta_L(l)&=1.366\cdot10^{-2}+5.811\cdot10^{-4}(l-1/2)\\
  \beta_R(l)&=1.366\cdot10^{-2}-5.811\cdot10^{-4} (l-1/2)~.
\end{align*}
From \cite{mcmanus91}, the observed fractions of left-handed offspring $\sigma_{XY}$ are  
\begin{align*}
  \sigma_{RR}(l^*)&=0.095~,\\
  \sigma_{RL}(l^*)&=0.195~,\\
  \sigma_{LL}(l^*)&=0.261~.
\end{align*}
We expect these parameters to be functions of $l$, but all data is drawn from modern societies where the fraction left-handed is $l=l^*$. Fortunately, using symmetry arguments, we can obtain additional points:
\begin{align*}
  \sigma_{RR}(1-l^*)&=1-0.261~,\\
  \sigma_{RL}(1-l^*)&=1-0.195~,\\
  \sigma_{LL}(1-l^*)&=1-0.095~.
\end{align*} 

In a population of uniform handedness, one might expect all offspring to inherit the same handedness as their parents.  However, in practice the situation is more complex.  Monozygotic (identical) twins often possess discordant handedness \cite{mcmanus91}.  Thus, handedness cannot be fully determined by genotype.  To account for this, most genetic models introduce a random component that partially determines handedness.  With that motivation, we define $\epsilon_{XY}$ to be the probability due to chance that parents with handedness $XY$ produce left-handed offspring in a population consisting entirely of right-handers. We then obtain: 
\begin{align*}
  \sigma_{RR}(0)&=\epsilon_{RR} &\sigma_{RR}(1)&=1-\epsilon_{LL} \\
  \sigma_{RL}(0)&=\epsilon_{RL} &\sigma_{RL}(1)&=1-\epsilon_{RL} \\
  \sigma_{LL}(0)&=\epsilon_{LL} &\sigma_{LL}(1)&=1-\epsilon_{RR}~.
\end{align*}  
It is unclear exactly what values are appropriate for $\epsilon_{XY}$ since no isolated human population consisting entirely of left- or right-handers exists.  If $\epsilon_{XY}=0$, then equilibria at $l=0,l^*,1/2,1-l^*,1$ appear and those at $l=l^*$ and $l=1-l^*$ are unstable.  This is inconsistent with the observed stable fixed point.  We therefore assume $\epsilon_{XY}$ must satisfy $0<\epsilon_{XY}<\sigma_{XY}(l^*)$. 

For given $\epsilon_{XY}$ values, we fit a cubic polynomial to the 4 known points $\sigma_{XY}(l)$ (known at $l=0,l^*,1-l^*,1$) to obtain smooth approximate functions $\sigma_{XY}(l)$.  The resulting dynamical system governed by equation~\eqref{phenotypic_model} has either 3 or 5 fixed points:  ($l^*,\frac{1}{2},1-l^*$) or ($l^*,l_2,\frac{1}{2},1-l_2,1-l^*$) where $l^*$ and $1-l^*$ are stable.  For example, if we set $\epsilon_{RR}=\epsilon_{RL}=\epsilon_{LL}=0.02$, we see an unstable fixed point at $l=1/2$ in addition to the expected stable fixed points at $l=l^*$ and $l=1-l^*$.  

Up to this point, we have treated $l^*$ as an unknown parameter.  In practice, however, the fraction of the population that is left-handed can be measured.   Estimates for the value of this parameter depend on the precise method of measurement and definition of left-handedness, but it is generally agreed that this fraction is close to 10\% \cite{frayer11}.  Equation~\eqref{phenotypic_model} allows us to compute an independent prediction for $l^*$ using only the birth rates and the phenotype ratios of offspring given above (note: the predicted $l^*$ does not depend on the choice of $\epsilon$ although the stability of the fixed point does). This results in a predicted percent left-handed of $11.78\%$, consistent with the measured value.

\begin{figure}[h] 
\psfrag{x1val}[][]{$0$}
\psfrag{x2val}[][]{$1/2$}
\psfrag{x3val}[][]{$1$}
\psfrag{y1val}[r][r]{$-3 \times 10^{-4}$}
\psfrag{y2val}[r][r]{$0$}
\psfrag{y3val}[r][r]{$ 3 \times 10^{-4}$}
\psfrag{insetxlabel}[b][b][2][0]{\raisebox{-1cm}{\hspace{1cm}$l$}}  
\psfrag{insetylabel}[][][2][-90]{$\frac{dl}{dt}$}  
\begin{center}  
\includegraphics[width=84mm]{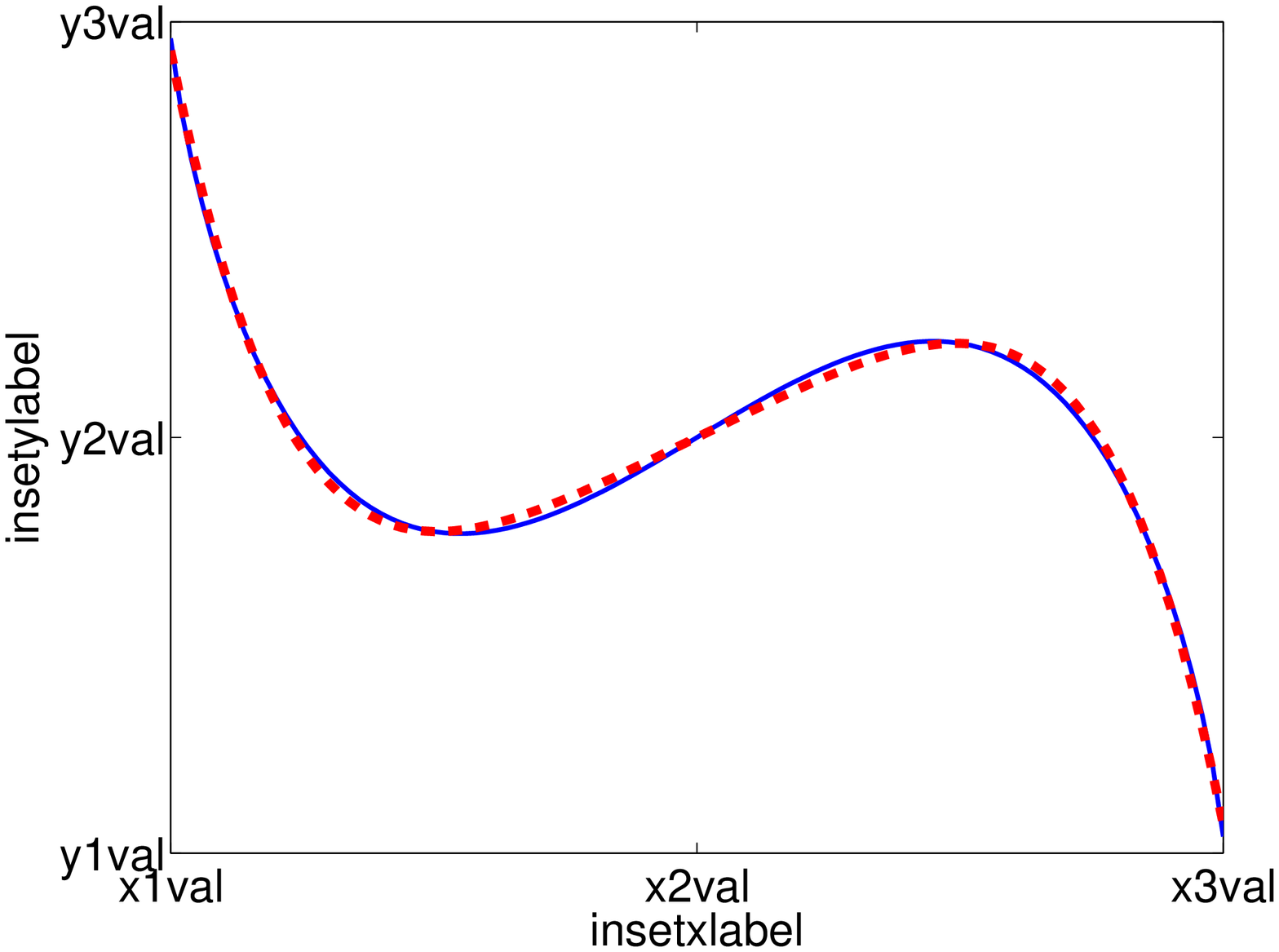}
\caption{Comparison of probabilistic and phenotypic models.  Solid blue line: the function $dl / dt$ [\%/yr] generated using generic sigmoid $P_{RL}$ in the probabilistic model from equation~\eqref{ode_v2}~of the main text.  Dashed red line: the function $dl/dt$ [\% / yr] implied by the phenotypic model from equation~\eqref{phenotypic_model}.}  \label{fig:ldotplot}
\end{center}
\end{figure}

By presenting this model of phenotype evolution, we wish to emphasize the generality of the probabilistic model presented in the main text.  For appropriate choices of functions $P_{RL}$, equation~\eqref{ode_v2}~in the main text can be made to agree nearly identically with model \ref{phenotypic_model} above, as demonstrated in figure~\ref{fig:ldotplot}.  

\section{Relating fitness functions to probabilistic transition rates}
\label{sec:Relating fitness functions to probabilistic transition rates}

Often in models of population dynamics, changes in the makeup of a population are related directly to the comparative fitness of the individuals within that population.  For example, Ghirlanda et al.~define a fitness function for lateralized individuals in terms of two component fitness functions, an antagonistic function and a synergistic function.  They then argue that equilibrium is obtained when the fitness of left- and right-handed individuals are equal \cite{ghirlanda09}.

In this paper, however, we employ a different approach.  We argue that segments of a population will switch handedness over long time scales according at a rate determined by a probabilistic function.  Thus equilibrium is obtained when the overall transition rates between left- and right-handed individuals balance.  These probabilistic transition rates should be related to the comparative fitness of the members of the population. 

We now determine the relationship between fitness functions $f_L(l)$, $f_R(l)$ and probabilistic transition rates $P_{RL}(l)$, $P_{LR}(l)$.  Recall that $P_{RL}(l)$ represents the probabilistic transition rate from right to left.  Clearly, $P_{RL}(l)$ must be non-negative. Also, it should reach a maximum (minimum) when the difference in fitness between left- and right-handers is at a maximum (minimum).  The simplest example of such a function is $P_{RL}(l) = A (f_L(l)-f_R(l)) + B$, where $A, B > 0$ are constants guaranteeing that $P_{RL}(l)$ remains positive.  By symmetry, $f_L(l)=f_R(1-l)$ and $P_{RL}(l)=P_{LR}(1-l)$.  Thus $P_{LR}(l)=A(f_R(l)-f_L(l))+B$.  Transition rates defined in this way satisfy the symmetry relation $P_{RL}(l)+P_{RL}(1-l)=2P_{RL}(1/2)$ and, as such, are sigmoidal for many different types of fitness functions.  

We should note that in this model, the probability $P_{RL}(l)\Delta t$ that a given individual switches from right to left within time $\Delta t$ is non-zero even when $f_R(l)>f_L(l)$. In such a situation, the probability that a given left-hander switches is $P_{LR}(l) \Delta t > P_{RL}(l) \Delta t$.  However, if right-handers are much more prevalent than left-handers, the total number of switches from right to left may still outweigh the number from left to right.  In other words, it is possible that $l P_{LR}(l) \Delta t < r P_{RL}(l) \Delta t$.  This would cause the fraction left-handed to increase despite right-handers having a higher fitness. 

Also, if for some $l_0$, $f_R(l_0)=f_L(l_0)$, then $P_{RL}(l_0) = P_{LR}(l_0) = B$ and $\left.\frac{dl}{dt}\right|_{l_0} = (1-l_0)B - l_0(B) = (1 - 2l_0)B$.  So if $l_0 < 1/2$, the fraction left-handed will increase even when the fitness is equal for left- and right-handers.  In our model, having equal fitnesses does not necessarily lead to equilibrium.

\begin{figure}[ht!]
\psfrag{x1val}[][]{$0$}
\psfrag{x2val}[][]{$1/2$}
\psfrag{x3val}[][]{$1$}
\psfrag{y1val}[r][r]{$-3 \times 10^{-4}$}
\psfrag{y2val}[r][r]{$0$}
\psfrag{y3val}[r][r]{$ 3 \times 10^{-4}$}
\psfrag{insetxlabel}[b][b][2][0]{\hspace{1cm}$l$}
\psfrag{insetylabel}[b][b][2][-90]{$\frac{dl}{dt}$}
\begin{center}  
\includegraphics[width=84mm]{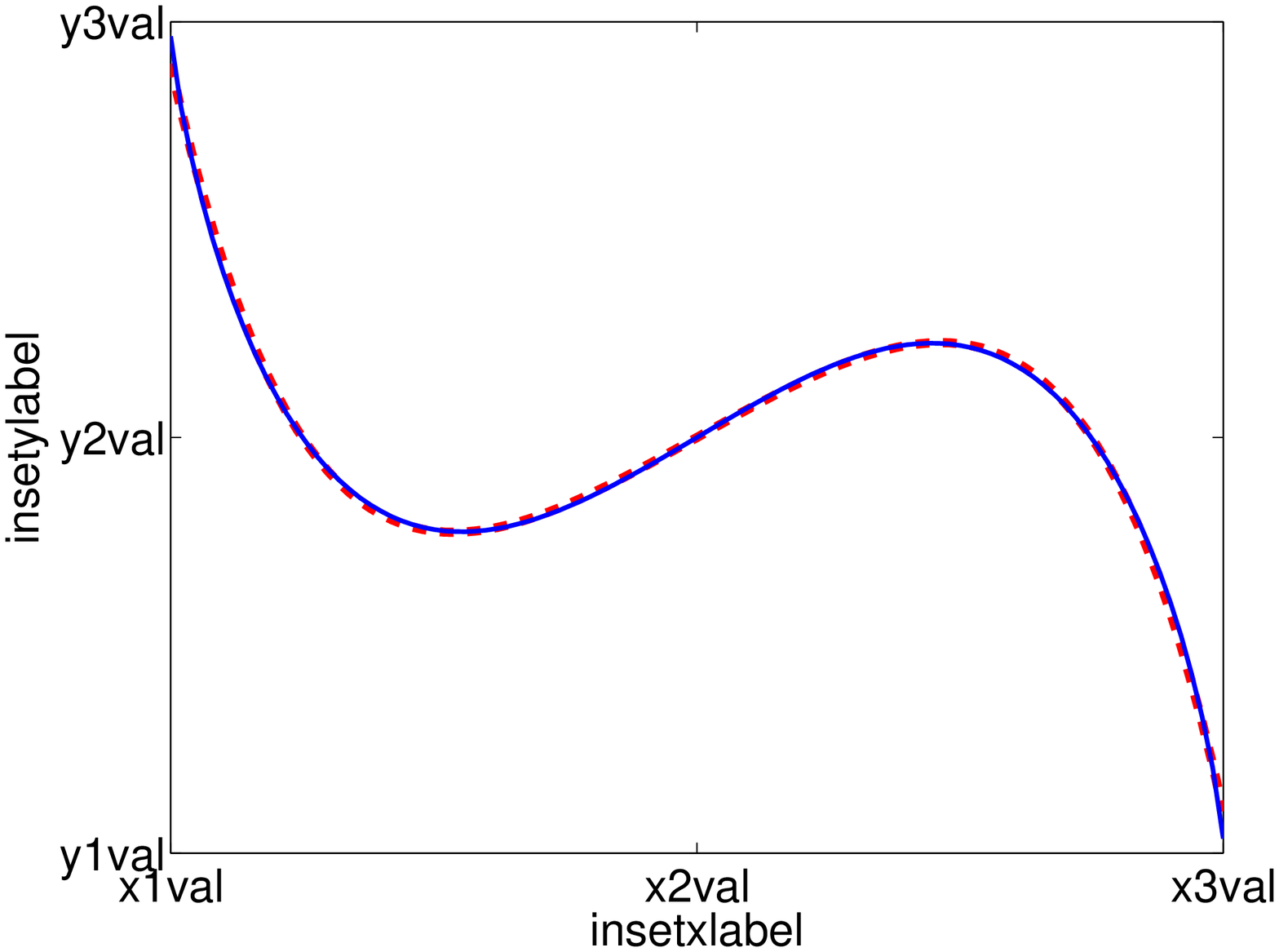}
\caption{Comparison of probabilistic and fitness-function models.  Solid blue line: the function $dl /dt$ [\%/yr] generated using generic sigmoid $P_{RL}$ in the probabilistic model from equation~\eqref{ode_v2}~of the main text.  Dashed red line: the function $dl/dt$ [\%/yr] generated using $P_{RL}$ implied by fitness functions proposed by Ghirlanda et al.~\cite{ghirlanda09}.} \label{fig:ldotplot_fitness}
\end{center}
\end{figure}

Despite the differences between our formulation and a fitness-based formulation, the predictions are similar. If we use the fitness functions $f_L(l)=(1-c)e^{-k_al}+c(1-e^{-k_sl})$ similar to those proposed by Ghirlanda et al.~we can generate $P_{RL}(l)$ and $P_{LR}(l)$ (which are sigmoid as expected).  In figure~\ref{fig:ldotplot_fitness} we plot the resulting function $\frac{dl(t)}{dt}$ for appropriate parameter values (such as $A=0.0144$, $B=0.0047$, $c=0.6964$, $k_a=3.7745$, $k_s=1.9974$) and observe a graph very similar to figure~\ref{fig:ldotplot}.  

\section{Derivation of athletic selection model}
\label{sec:Derivation of athletic selection model}

Professional sports are artificial systems that involve varying degrees of competitive and cooperative activities.  Their participants undergo a selection process through tryouts that is in some ways analogous to natural selection.  Additionally, handedness data for professional athletes is widely available.  Thus, athletics provide an ideal opportunity to test whether our model's predictions are consistent with data from selective systems.

There is a fundamental difference between selection in professional sports and natural selection.  In natural selection, the distribution of a trait within a population changes in response to selection pressure, modifying the gene pool.  Professional athletes, however, represent only a small segment of the much larger human population.  Changes in the distribution of a trait among professional athletes are unlikely to influence the gene pool in the human population; furthermore, most professional sports have not existed for the time scales required to significantly modify the gene pool.  As a result, the population of professional athletes must draw new members from a pool that consists of about 90\% right-handers.  Thus, direct comparison of our model to sports data is not possible: instead, we must account for this more complex selection process in order to make predictions that \textit{are} applicable to real-world sports. 

To begin, we define $l^*$ to be the fixed point predicted by the probabilistic model \eqref{ode_v2}~described in the main text. This represents the ideal equilibrium distribution of left-handedness in a hypothetical world where all interaction occurs through the sport under consideration. We assume that skill is normally distributed throughout the population with mean $\mu=0$ and standard deviation $\sigma$.  Because left-handedness is relatively rare, this trait should provide a competitive advantage in sports involving direct physical confrontation. Let $l$ represent the fraction left-handed within a sport.  When $l$ deviates from $l^*$, the sport is not at its ideal equilibrium state, and left-handers must experience a shift in skill $\Delta s$.  We assume that professional sports operate efficiently, that is, they select players exclusively according to skill level.  Then the distributions of skill among left- and right-handers satisfy
\begin{equation*}
    \label{pdf_righties}
   p_R\left(s\right)=\frac{1}{\sqrt{2\pi\sigma^2}}e^{-\frac{s^2}{2\sigma^2}}
\end{equation*}
\begin{equation*}
    \label{pdf_lefties}
   p_L\left(s\right)=\frac{1}{\sqrt{2\pi\sigma^2}}e^{-\frac{\left(s-\Delta s\right)^2}{2\sigma^2}}
\end{equation*}  
where  
\begin{align*}
\Delta s &= k \left(l^*-l\right)~.
\end{align*}
In this formulation, when the ideal equilibrium fraction of left-handedness is achieved, their is no advantage to possessing either handedness and the individuals are selected according their intrinsic skill.

In most sports, individuals must undergo a tryout in order to demonstrate sufficient skill for participation. As a result only a fraction of the total population is allowed to participate in the sport at a given level of competition.  We define the fraction selected $\psi = n/N$, where $n$ is the number of individuals selected and $N$ is the size of the total population. $\psi$ will determine a minimum skill cutoff $s_c$ for participation according to the relation  
\[
  \psi = \lbg \int_{s_c}^{\infty}p_L\left(s\right)ds + \left(1-\lbg\right) \int_{s_c}^{\infty}p_R\left(s\right) ds~,
\]
where $\lbg$ represents the background rate of left-handedness ($\approx 1/10$).  We can simplify this expression by normalizing the various parameters by the standard deviation: we set $\schat \equiv s_c / \sqrt{2\sigma^2}$, $\Dshat \equiv \Delta s / \sqrt{2\sigma^2}$, and $\khat \equiv k / \sqrt{2\sigma^2}$ to get
\begin{equation}  \label{psieq}
  \psi = \frac{1}{2} \lbg \erfc \left(\schat - \Dshat \right) + \frac{1}{2}(1-\lbg) \erfc \left(\schat\right) ~.
\end{equation}
In this formulation, the fraction left-handed among the individuals selected will be represented by the first term on the right-hand side of equation~\eqref{psieq} divided by the entire expression,
\[
  \lsel = \frac{\lbg}{2\psi} \erfc \left( \schat-\Dshat \right).
\]

After many tryouts, the fraction left-handed within a sport will stabilize with the $\lsel=l$.  This equilibrium fraction represents the observed fraction left-handed among professional athletes, thus we call it $\lpro$.  So, the equilibrium state is implicitly determined by 
\begin{equation}  \label{lpro_eqs}
  \lpro = \frac{\lbg}{2\psi} \erfc \left[\schat - \khat \left( l^*-\lpro \right)\right]~.
\end{equation}
This model has the following properties: 
\begin{itemize}
\item
If $l^*=\lbg$, then $\lpro=\lbg$. 
\item
If the sport is not selective at all, in the limit $\schat \to -\infty$ ($\psi \to 1$), $\lpro \to \lbg$ as this means that selection is independent of skill. 
\item
As the sport becomes infinitely selective, $\schat \to \infty$ ($\psi \to 0$), $\lpro \to l^*$. In other words, the ideal equilibrium fraction is achieved when the sport is infinitely selective. (To see this, assume that $\lsel = l^* + \delta$ and $l = l^* - \delta$.  Expand in a Taylor series about $\delta = 0$ and take the limit $\schat \to \infty$ to find that $\delta \to 0$.  Thus $\lsel \to l^*$ and $l \to l^*$, so $\lpro = l^*$.)
\end{itemize}

Using this model, we employed numerical techniques to compute the solutions $\lpro$ for a variety of sports, and then compared these results to the observed fractions left-handed as seen in figure~2.  

This model can also be extended to examine how the distribution of handedness varies within a single sport.  If left-handedness is a desirable trait (that is, it provides a skill advantage at equilibrium), then we expect that it should be very prevalent among the most skilled individuals due to the selection mechanism.  To see this, we consider the case of baseball.  It is clear that in baseball, $l^*=\frac{1}{2}$ since the sport involves primarily competitive interactions (the observed $\lpro \approx 0.3$). Thus, left-handedness is a desirable trait for potential professionals as it will provide a particular skill advantage in batting.  At the professional level, most hitters face the same set of pitchers and compete indirectly with one-another for roster spots.  They should therefore be expected to experience the same skill advantage due to handedness.  In other words, $\Dshat = \khat(l^*-\lpro)$ is a constant (in sports like boxing, however, where individuals compete more frequently with others near their own rank, the skill advantage would be a rank-dependent function $r$, $\Dshat_r = \khat(l^*-l_r)$).  Ranking the hitters by skill, we observe that the fraction left-handed above rank $r$ should satisfy 
\begin{equation*}  \label{lreq}
  l_r = \frac{\lbg N}{2r} \erfc\left(\srhat - \Dshat \right)~,
\end{equation*} 
where $\srhat$ satisfies
\begin{equation*}  \label{req}
  r = \frac{1}{2} \lbg N \erfc \left(\srhat - \Dshat \right) + \frac{1}{2}(1-\lbg) N \erfc \left( \srhat \right)~.
\end{equation*}

Using this result, we plotted the predicted fraction left-handed as a function of rank as seen in figure~3.  This model predicts a non-trivial shape for the distribution of handedness within baseball that is consistent with the observed distribution.  This is a strong indication that this athletic selection model provides a good mathematical approximation for the tryout-based selective mechanism within professional sports.

\section{Data summary}
\label{sec:Data summary}
\begin{sidewaystable}
\small{
\begin{tabularx}{8in}{|p{2.0cm}|X|>{\centering}p{1.2cm}|>{\centering}p{0.8cm}|>{\centering}p{1.5cm}|>{\centering}p{1.5cm}|>{\centering}p{1.8cm}|p{2.0cm}|}
\hline 
\textbf{Sport} & \textbf{Data source} & \textbf{Total participants} & \textbf{Total pros} & \textbf{Observed fraction left-handed} & \textbf{Predicted fraction left-handed} & \textbf{Method} & \textbf{URL(s)}\\
\hline \hline
Baseball & Major League Baseball players 1871 to 2009, Lahman Baseball Database & 2629000 & 1200 & 0.329 & 0.318 & Handedness listed & \urlf{baseball1.com}\\
\hline
Boxing & BoxRec, top 200 ranked boxers by class: Heavyweight, Cruiserweight, Middleweight, Welterweight, Featherweight,  Bantamweight,  and Flyweight, September 2010 & 587000 & 1056 & 0.201 & 0.304 & Handedness listed & \urlf{boxrec.com}\\
\hline
Fencing, Men & Federation Internationale D'Escrime, top 25 ranked for each weapon: Epee, Saber and Foil, 2010 & 1462 & 75 & 0.200 & 0.247 & Handedness listed & \urlf{fie.ch}\\
\hline
Fencing, Women & Federation Internationale D'Escrime, top 25 ranked for each weapon: Epee,  Saber and Foil, 2011 & 1123 & 75 & 0.253 & 0.240 & Handedness listed & \urlf{fie.ch}\\
\hline
Football, Quarterbacks & National Football League, active players, preseason 2010 & 75071 & 120 & 0.067 & 0.064 & Google image search & \urlf{nfl.com} \\
\hline
Golf, Men & PGA TOUR, top earners 2009 & 4600000 & 120 & 0.042 & 0.059 & PGA Tour profile images & \urlf{pgatour.com}\\
\hline
Golf, Women & Ladies Professional Golf Association, top earners 2010 & 1000000 & 100 & 0.000 & 0.060 & LPGA Tour profile images & \urlf{lpga.com}\\
\hline
Hockey, Defensemen and Forwards & National Hockey League, active players,  preseason 2010 & 309971 & 601 & 0.323 & 0.303 & Shot side listed & \urlf{espn.go.com/nhl}\\
\hline
Hockey, Left Wings & National Hockey League, active players,  preseason 2010 & 80458 & 156 & 0.090 & 0.065 & Shot side listed & \urlf{espn.go.com/nhl}\\
\hline
Hockey, Right Wings & National Hockey League, active players,  preseason 2010 & 91805 & 178 & 0.742 & 0.613 & 
Shot side listed & \urlf{espn.go.com/nhl}\\
\hline
Table Tennis, Men & International Table Tennis Federation, top ranked players, November 2010 & 1586628 & 100 & 0.280 & 0.333 & Ranking photo & \urlf{ittf.com}\\
\hline
Table Tennis, Women & International Table Tennis Federation, top ranked players, November 2010 & 654372 & 100 & 0.200 & 0.327 & Ranking photo & \urlf{ittf.com} \\
\hline
\end{tabularx}
\caption{Summary of data sources used in preparing figure~2.}
\label{datatable}
}  
\end{sidewaystable}

Data used in generation figure~2 came from a variety of sources.  The total number of participants came from surveys conducted by the Sporting Goods Manufacturers Association in 2009 \cite{SGMA}, except for men's and women's fencing, where participant numbers were extrapolated from data published by the National Federation of State High School Associations \cite{highschooldata}.  The number of professional players came from listings of top-rated players (the only ones for which handedness was readily available) at the internet URLs indicated in Table \ref{datatable}, with the exception of baseball, football, and hockey, where numbers are absolute totals.

When handedness was not available in tabulated form, it was evaluated based on public photos of players in action.

In Table \ref{datatable}, the predictions for the fraction left-handed were generated using an estimate of the ideal equilibrium $l^*$ for each sport.  The appropriate value for $l^*$ depends primarily on the degree of cooperation $c$ for the sport.  This parameter is difficult to estimate in sports that possess clear cooperative and competitive elements.  However, in order to observe fixed points other than $l^*=1/2$, $c$ must exceed a threshold that appears to be relatively high for the types of transition rates considered in this paper (See figure \ref{fig1}).  So, we assumed that $l^*=1/2$ for sports primarily involving direct confrontations: baseball (batters vs. pitchers), boxing, fencing, table tennis, hockey (defensemen and forwards).  

Some sports (or particular positions within sports), however, possess highly lateralized equipment, positioning or strategy.  For these sports, it is ideal for all individuals to possess the same handedness; so, the minority handedness will be selected against.  For example, in football, blocking schemes are often designed to protect a quarterback's blind side.  As a result, it is beneficial for all quarterbacks on the roster to possess the same handedness in order to minimize variations of the offensive sets.  Consequently, we assume that for quarterbacks in football, golfers, and left and right wings in hockey the value of $c \approx 1$, i.e., $l^* = 0 \textrm{ or } 1$.

\section{Parameter sensitivity analysis for probabilistic model}
\label{sec:Parameter sensitivity analysis for probabilistic model}   

In the probabilistic model, there are two unknown functions $\PRLcoop(l)$ and  $\PRLcomp(l)$.  While general properties of these functions such as monotonicity are known, the appropriate form for these functions is unknown and is difficult to determine from data.  The generic sigmoid functions 
\begin{subequations}  \label{PRL_sigmoid}
\begin{align}  
  \PRLcomp(l) &= \left[ 1+e^{k_1 \frac{1-2l}{l(1-l)}} \right]^{-1}  \label{PRL_sigmoid_comp} \\  
  \PRLcoop(l) &= \left[ 1+e^{-k_2 \frac{1-2l}{l(1-l)}} \right]^{-1}~,  \label{PRL_sigmoid_coop}
\end{align}
\end{subequations}
satisfy restrictions on $P_{RL}$ for $k_1, k_2 > \sqrt{3} / 2$ and capture the essential fixed point behaviour ($k_1,k_2$ set the steepness of the curves).  Unfortunately, these equations introduce two new parameters that may alter the dynamics.   To examine the sensitivity of the model to these parameters, we assumed $\lstar$ was the fixed point of the system.  We then computed the partial derivatives of $\lstar$ with respect to each parameter.  In the vicinity of $\lstar = 1/10$, the observed ratio of left-handers in human populations, and in the range of $k_1 \in \left[\frac{\sqrt{3}}{2},\infty\right)$, $k_2 \in \left[\frac{\sqrt{3}}{2},\infty\right)$ and $c\in(0,1)$, we found that
\[
\frac{\left|\frac{\partial \lstar}{\partial c}\right|}{\left|\frac{\partial \lstar}{\partial k_1}\right|} \geq O(10^2) \textrm{ and } \frac{\left|\frac{\partial \lstar}{\partial c}\right|}{\left|\frac{\partial \lstar}{\partial k_2} \right|} \geq O(10^2)~.
\]

Thus for a wide range of parameter values, $\left|\frac{\partial \lstar}{\partial c}\right|\gg\left|\frac{\partial \lstar}{\partial k_2}\right|$ and $\left|\frac{\partial \lstar}{\partial c}\right|\gg\left|\frac{\partial \lstar}{\partial k_1}\right|$ near the fixed point $\lstar = \frac{1}{10}$.  In other words, the location of the fixed point is more sensitive to $c$ than $k_1$ and $k_2$ by several orders of magnitude for physically allowable $k$-values.  Therefore we are justified in ignoring the effects of individual choices of $k_1$ and $k_2$ in order to focus on the effects of the choice of $c$.  We believe these results are robust for various different sigmoid functions $\PRLcomp, \PRLcoop$.


\end{document}